\RequirePackage{fix-cm}

\documentclass[pdftex,twocolumn,epjc3]{svjour3} 
\smartqed  % flush right qed marks, e.g. at end of proof
\usepackage{graphicx}
\usepackage{amsmath}
\usepackage{color}

\usepackage{epstopdf}

\RequirePackage{fix-cm}
\RequirePackage{graphicx}
\RequirePackage{mathptmx}      % use Times fonts if available on your TeX system
\RequirePackage{flushend}
\RequirePackage[numbers,sort&compress]{natbib}
\RequirePackage[colorlinks,citecolor=blue,urlcolor=blue,linkcolor=blue]{hyperref}

\begin{document}

\title{DEPFET detectors for direct detection of MeV Dark Matter particles}
%\thanks{Grants or other notes
%about the article that should go on the front page should be
%placed here. General acknowledgments should be placed at the end of the article.}

%\subtitle{Do you have a subtitle?\\ If so, write it here}

%\titlerunning{Short form of title}        % if too long for running head

\author{A. B{\"{a}}hr\thanksref{addr1}
\and H. Kluck\thanksref{addr2,addr3}
\and J. Ninkovic\thanksref{addr1}
\and J. Schieck\thanksref{e1,addr2,addr3}
\and J. Treis\thanksref{addr1}}

\thankstext{e1}{e-mail: Jochen.Schieck@oeaw.ac.at}

%\authorrunning{Short form of author list} % if too long for running head

\institute{Max-Planck-Gesellschaft Halbleiterlabor, Otto Hahn Ring 6, D-81739 M\"unchen, Germany\label{addr1}
          \and
          Institut f\"ur Hochenergiephysik der \"Osterreichischen Akademie der Wissenschaften, Nikolsdorfer Gasse 18, A-1050 Wien, Austria\label{addr2}
          \and
         Atominstitut, Technische Universit\"at Wien, Stadionallee 2,  A-1020 Wien, Austria\label{addr3}
}

\date{Received: date / Accepted: date}
% The correct dates will be entered by the editor

\maketitle

\begin{abstract}
The existence of dark matter is undisputed, while the nature of it is still unknown. 
Explaining dark matter with the existence of a new unobserved particle is among
the most promising possible solutions. Recently dark matter candidates in the 
MeV mass region received more and more interest. In comparison to the
mass region between a few GeV to several TeV, this region is experimentally largely 
unexplored. We discuss the application of a RNDR DEPFET semiconductor detector 
for direct searches for dark matter in the MeV mass region. We present
the working principle of the RNDR DEPFET devices and review the performance obtained 
by previously performed prototype measurements. The future potential of the technology
as dark matter detector is discussed and the sensitivity for MeV dark matter detection 
with  RNDR DEPFET sensors is presented. 
Under the assumption of six background events in the region of interest and an 
exposure of one kg$\cdot$y a sensitivity of about $\overline{\sigma}_{e} = 10^{-41}\,\mathrm{cm^2}$ for 
dark matter particles with a mass of 10 MeV can be reached. 

\end{abstract}

\section{Introduction}
Several independent measurements clearly point towards the existence of dark matter. The nature
of dark matter is still not understood and is among the biggest outstanding problems of modern 
physics~\cite{Bertone:2004pz}. A well motivated solution to this problem is the existence of a 
new particle candidate, which interacts at most
weakly with standard model particles. The possible mass range of this particle candidate, as well as the possible 
interaction
strength with ordinary matter, spans several orders of magnitude~\cite{Baer:2014eja}. Recently 
several theoretical studies focus on possible dark matter candidates in the MeV mass region, 
below the mass scale of weakly interacting massive 
particles~\cite{Kaplan:1991ah,Petraki:2013wwa,Kaplan:2009ag,Boehm:2003hm,Boehm:2002yz,Feng:2008ya}. This mass region is experimentally
less explored and opens a large space for undiscovered dark matter candidates. \\
Direct detection experiments search for relic dark matter particles by looking for 
elastic scatterings between a dark matter candidate and a nucleus. 
The energy deposited in the scattering processes, the nuclear recoil-energy, can be measured by the experiment.
By using simple kinematic relations the mass of the dark matter particle can be inferred from the recoil energy. 
The sensitivity towards low mass dark matter particles is determined by the detection threshold 
for nuclear recoils. Dark matter candidates with masses below $100 \,\mathrm{MeV}$ lead to a nuclear recoil 
as low as a few eV and are therefore below the threshold of direct dark matter detection experiments. 
The search for light dark matter particles via scattering with an electron opens the opportunity
to extend the reach towards even smaller masses, down to a few MeV. However, the theoretical
prediction of the dark matter-electron scattering rate is more complex, compared to
the nuclear scattering. In this paper we study the possibility to measure dark matter-electron
scattering using a silicon based detector. In solid state detectors electrons are bound to the 
nucleus and can no longer being considered as free particles. Electrons are not at rest and
the typical speed is greater compared to the average speed of the dark matter particle, leading 
to a different kinematics of the process. In addition the complicated 
electronic structure of the semiconductor makes the calculation of the scattering rate more
complicated. This topic has been discussed in detail in the literature, e.g. 
in~\cite{Essig:2011nj,Essig:2015cda,Graham:2012su,Lee:2015qva}, and is only summarised here. \\
A semiconductor detector based on the DEPFET principle (DEpleted P-channel Field Effect Transistor) with
repetitive non-destructive readout (RNDR)~\cite{Kemmer:1986vh} offers the possibility to perform a low-noise measurement 
of the ionisation signal originating from a dark matter-electron inelastic scattering process, down 
to a single electron. The excellent noise performance for the ionisation signal
is reached by repeating the measurement in a statistically independent way.
With the average ionisation energy for a single electron of a few eV the detector performance 
can be transformed to a sensitivity for dark matter masses down to a few MeV. \\
We briefly review the detection of MeV dark matter with semiconductor targets in section~\ref{SiliconDarkMatterDetection},
the RNDR DEPFET detector principle and the expected detector performance is discussed in section~\ref{DEPFET}
and in section~\ref{ExpSensitivity} we present the expected sensitivity for MeV direct dark matter detection.
In section~\ref{Summary} we summarise the potential of RNDR DEPFET detectors for direct dark matter detection.
\section{Detection of MeV dark matter by dark matter-electron scattering}
\label{SiliconDarkMatterDetection}
\begin{figure}[]
\begin{center}
        \includegraphics[width=0.450\textwidth]{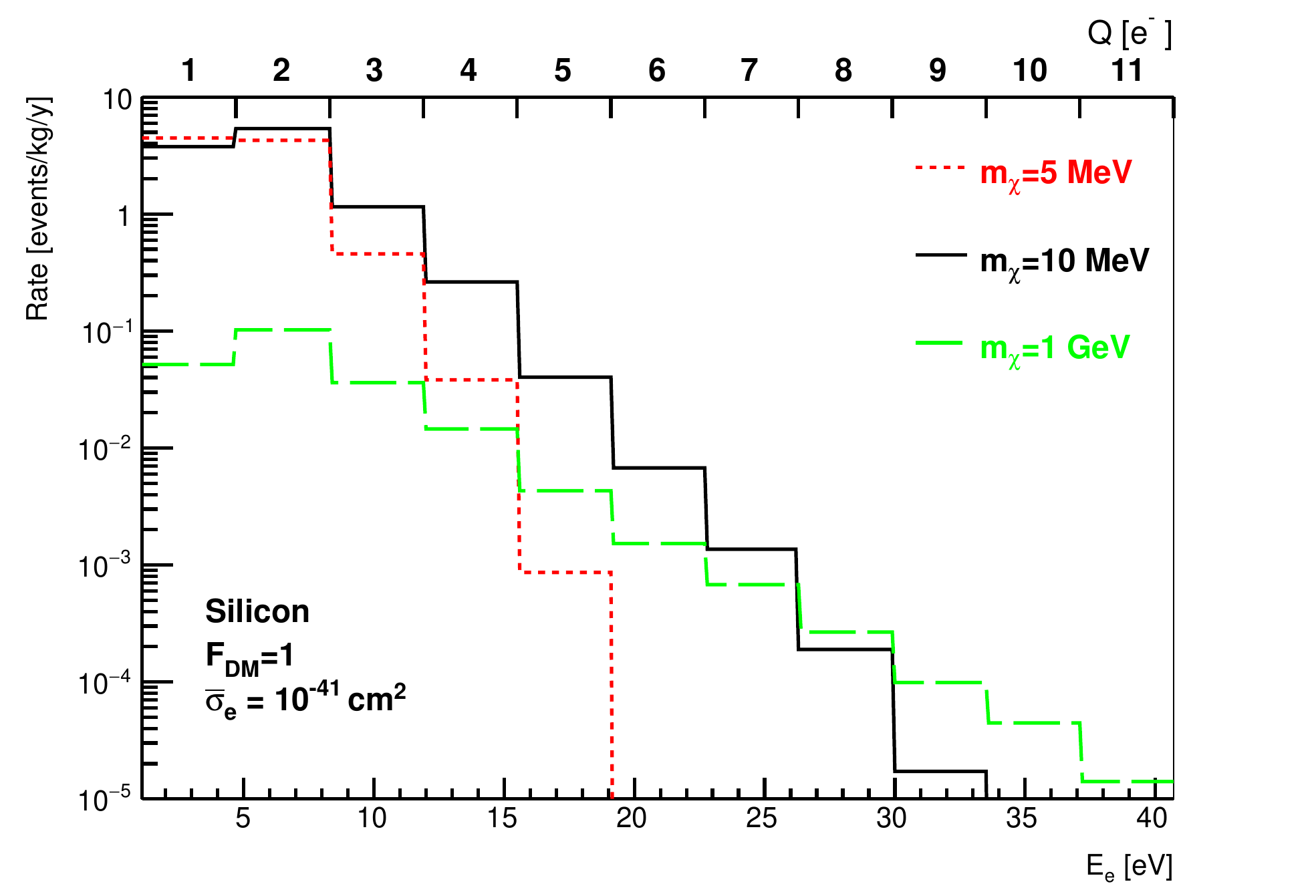}
\caption{The expected recoil spectrum for a non-relativistic dark matter elastic 
cross-section of $\overline{\sigma}_\mathrm{e} = 10^{-41}\,\mathrm{cm^2}$ as a function
of the deposited energy $E_\mathrm{e}$ and the ionization $Q$.
The rate is
shown for incoming dark matter particles with a mass of 
$1\,\mathrm{GeV}$ (green, dashed), $10\,\mathrm{MeV}$ (black, solid) and $5\,\mathrm{MeV}$ (red, dotted). We assume
standard astrophysical assumptions for the dark matter density and velocity 
distribution~\cite{DMVelocity}. The form factor $\mathrm{F_{DM}}$ is set to one. }
\label{fig:ER_Si}
\end{center}
\end{figure}

The process of dark matter-electron scattering is derived and discussed in references
~\cite{Essig:2011nj,Essig:2015cda,Graham:2012su,Lee:2015qva}.
In this section we summarise the key findings of~\cite{Essig:2015cda}, which are
necessary to discuss the expected sensitivity
for RNDR DEPFET dark matter detectors in section~\ref{ExpSensitivity}. The 
reader is referred to~\cite{Essig:2015cda} for a the complete derivation, in particular
about the crystal form factor of silicon, which contains relevant information about the
electron binding in the corresponding material.\\
The measurement of the recoil energy distribution from the dark matter scattering process, together 
with the expected velocity distribution of the dark matter gives an estimate of the mass of the 
incoming dark matter particle. For dark matter-nucleus scattering the mass can be 
derived by simple kinematic calculations and the deposited recoil energy is proportional to
$\frac{1}{m_\mathrm{N}}$, with $m_\mathrm{N}$ being the mass of the target
nucleus. A lighter target material therefore returns an increased average recoil energy,
which is experimentally easier to measure. 
The scattering between a dark matter particle and an electron is more complicated and 
requires a careful discussion. Compared to the dark matter-nucleus scattering no
simple interpretation of the scattering rate in terms of cross-section and dark matter mass is possible.
Two points are discussed in order to understand the relation between the
dark matter scattering rate and the underlying dark matter parameters: the
kinematic relation of the scattering process of MeV dark matter particles 
and the relevant binding effects of electrons in silicon. 
In a solid state device made of silicon electrons are bound and cannot be considered to be at rest.
The energy transferred to the electron $E_\mathrm{e}$ can be derived from  
a simple energy conservation relation $E_\mathrm{e} = - \Delta E_\mathrm{\chi}  - E_\mathrm{N}$~\cite{Essig:2015cda}, with $\Delta E_\mathrm{\chi}$
being the energy loss of the dark matter particle and $E_\mathrm{N}$ being the recoil energy of the whole atom.
Please note that the energy $E_\mathrm{e}$ is the total energy and only parts
of the energy is finally transferred as the kinetic energy of the electron, while the 
rest is needed to move the electron from the valence band to the conduction band.
We consider small energy transfers only and therefore 
the recoil energy of the atom, $E_\mathrm{N}$, can be safely set to zero.
The average velocity of the electron can be related to its binding energy, 
v$_\mathrm{e} \sim Z_\mathrm{eff} \, \alpha$, with $\alpha\approx 1/137$ being the fine-structure constant and the effective charge of the nucleus $Z_\mathrm{eff}$ being one for outer electrons.
The velocity v$_\mathrm{e}$ is large compared to 
the velocity of the incoming  dark matter particle, v$/c \sim 10^{-3}$. 
The average momentum transfer of the scattering process is therefore dominated 
by the momentum of the bound electron. This information, together with the energy 
conservation relation, can be used to show that the typical available momentum
transfer $q$ in MeV dark matter scatterings is enough to move electrons from
the valence band to the conduction band of silicon, 
with a band gap in the order of a few eV. \\
Parts of the energy transferred from the dark matter particle to the electron $E_\mathrm{e}$ 
is needed to move the electron from the valence band to the conductance band.
To predict the electron scattering rate the relevant electron 
binding effects for silicon need to be calculated. The calculation of a
dimensionless crystal form factor $f_\mathrm{crystal}(q,E_\mathrm{e})$ was performed 
for the first time in~\cite{Essig:2015cda} and can be considered as a 
key input to the prediction of the dark matter-electron scattering rate in silicon. 
The form factor calculation implies that the scattering processes with
larger $q$-values are suppressed compared to processes with 
low $q$, leading to a sensitivity increase towards low energy recoils. 
The differential recoil rate can be written as~\cite{Essig:2015cda}:
\begin{eqnarray}
\frac{dR}{d\,\ln E_\mathrm{e}}=\frac{\rho_\mathrm{\chi}}{m_\mathrm{\chi}}\, N_{\mathrm{cell}} \, \overline{\sigma}_\mathrm{e}\, \alpha \, \frac{m_\mathrm{e}^{2}}{\mu^{2}_\mathrm{\chi\,e}} \nonumber \\
\times \int d \ln q \bigg(\frac{E_\mathrm{e}}{q} \eta \big( v_\mathrm{min}(q,E_\mathrm{e}) \big)\bigg)\,  \nonumber \\ F_\mathrm{DM}(q)^{2}\, |f_\mathrm{crystal}(q,E_\mathrm{e})|^{2},
\label{Eq:Scattering}
\end{eqnarray}
with $\rho_{\chi}$ being the local dark matter density, $m_\mathrm{\chi}$ the mass of the dark matter particle, $N_{\mathrm{cell}}$ the number of unit cells in the target, $\overline{\sigma}_\mathrm{e}$
parametrizing the strength of the interaction, $m_\mathrm{e}$  the mass of the electron,  $\mu_\mathrm{\chi\,e}$ the reduced mass of the
dark matter-electron system and  $\eta \big( v_\mathrm{min}(q,E_\mathrm{e}))$ parametrizing the dark matter density profile. 
The dark matter form factor $F_\mathrm{DM}(q)$ parametrises the momentum dependence of the interaction. 
For $F_\mathrm{DM}(q)=1$ the interaction \linebreak strength $\overline{\sigma}_\mathrm{e}$ is reduced to a simple point like interaction.
$F_\mathrm{DM}(q)=(\alpha\, m_\mathrm{e}/q)$ corresponds to an electric dipole moment and $F_\mathrm{DM}(q)=(\alpha\, m_\mathrm{e}/q)^{2}$
corresponds to the exchange of a massless (or ultra-light) vector mediator. For our studies we choose the simplest  
momentum dependency and we set $F_\mathrm{DM}(q)=1$, as expected for a point-like interaction. \\
The energy deposited via the  dark matter scattering process $E_\mathrm{e}$ is converted to 
an average number of produced electrons $Q$, by setting the average ionisation
energy to $E_\mathrm{ion}=3.6\,\mathrm{eV}$ and the band-gap energy to $E_\mathrm{gap}=1.11\,\mathrm{eV}$.
The ionization $Q$ is given
by 
\begin{equation}
Q(E_\mathrm{e})= 1 + \mathrm{Int}[(E_\mathrm{e}-E_\mathrm{gap})/E_\mathrm{ion}],
\label{AvIonisation}
\end{equation}
~\cite{Essig:2015cda}.  The expected recoil rate as a function of deposited energy $E_\mathrm{e}$ is shown in
figure~\ref{fig:ER_Si}. For a $10\,\mathrm{MeV}$ dark matter particle about $65\%$ of all events generate   
at least two electrons in the detector.  The rate is calculated by using the publicly available QEdark 
code~\cite{Essig:2015cda}~\footnote{\tt http://ddldm.physics.sunysb.edu/ddlDM/}.
The expected sensitivity is presented after discussing the expected performance of the RNDR DEPFET device in 
section~\ref{ExpSensitivity} in terms of detected electrons.
\section{RNDR DEPFET sensors for direct dark matter detection}
\label{DEPFET}
\subsection{Concept of RNDR DEPFET devices}
\begin{figure}[]
\begin{center}
        \includegraphics[width=0.450\textwidth]{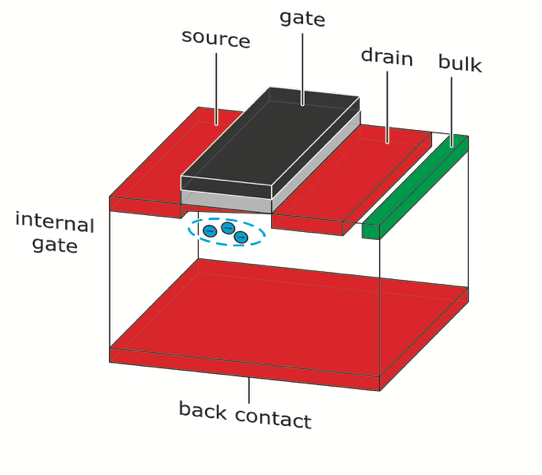}
\caption{Structure of a basic DEPFET cell.}
\label{fig:DEPFET_SingleCell}
\end{center}
\end{figure}

\begin{figure}[]
\begin{center}
   \includegraphics[width=0.35\textwidth]{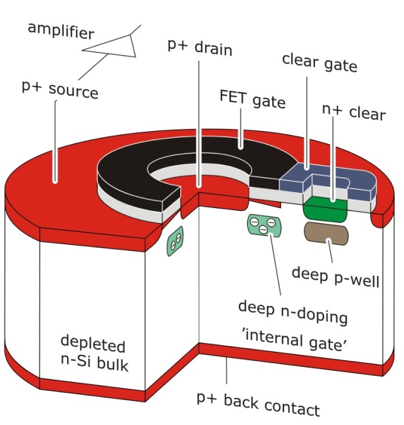}
  \includegraphics[width=0.35\textwidth]{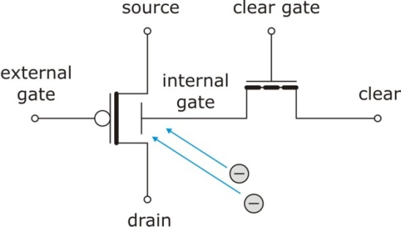}
\caption{Structure (top) and circuit representation (down) of a conventional spectroscopy-grade  DEPFET cell.}
\label{fig:DEPFET_CircularCell}
\end{center}
\end{figure}

\begin{figure}[]
\begin{center}
   \includegraphics[width=0.35\textwidth]{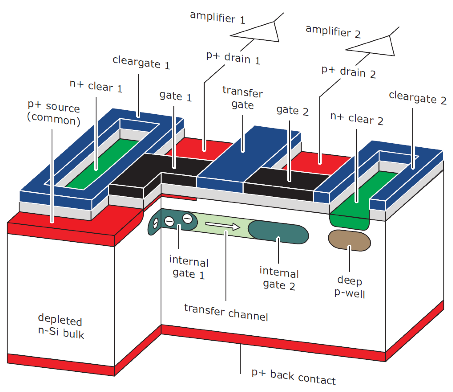}
  \includegraphics[width=0.35\textwidth]{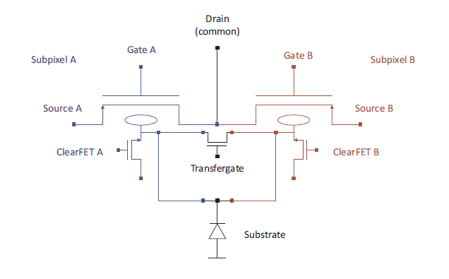}
\caption{Structure (top) and circuit representation (bottom) of  RNDR DEPFET {\it superpixel} consisting of two DEPFET {\it subpixels} with linear geometry.}
\label{fig:DEPFET_RNDR}
\end{center}
\end{figure}

The basic idea behind repetitive non-destructive readout\linebreak (RNDR) is to apply one of the most important implications
of the central limit theorem on the field of detectors. 
Any charge generated in the sensitive detector bulk is collected in the internal gates. 
Due to the excellent charge carrier lifetime, charge loss can be virtually excluded.
The  RMS noise of a single measurement is determined by the electronic noise of the transistor current measurement. 
By repetitively measuring the identical 
signal charge in a statistically independent way, the value resulting from the average of the individual 
measurements has a standard deviation of $\sigma_\mathrm{eff}=\frac{\sigma}{\sqrt{n}}$, with $\sigma$ being the RMS 
noise of a single measurement, and $n$ being the number of readings. In this
way, the standard deviation of the mean can be considered to be the effective noise of the measurement. \\
Devices based on the combined detector-amplifier structure DEPFET are applied for a variety 
of particle physics and astrophysical experiments~\cite{Ott:2016qhj,Marinas:2015vcq,Schieck:2013xta}. In their most 
simple form, they provide an active pixel sensor with pixel-individual charge storage and readout at high speed 
with very good signal-to-noise ratio (SNR). In addition, however, they provide 
an ideal platform to realise the RNDR principle for radiation detectors. \\
The simplest DEPFET cell~\cite{Kemmer:1986vh} consists of a P-channel FET integrated on a silicon bulk, which is fully depleted by means of sidewards depletion 
(see figure~\ref{fig:DEPFET_SingleCell}). By an additional deep-n implant directly below the gate, a potential minimum for electrons is created, which all bulk-generated 
electrons will drift to. In case a transistor current is present, their presence modulates the conductivity of the transistor channel, and this modulation is 
detected by appropriate subsequent electronics. Hereby, the potential minimum has the same effect on the channel as the external gate, and it is therefore also 
referred to as internal gate. High-accuracy measurements rely on correlated double-sampling (CDS) to determine the amount of charge. After an initial 
measurement of the transistor state, the charge is removed from the internal gate by an attached n-channel MOSFET, the ClearFET, and the transistor state is 
measured again with empty internal gate. The actual amount of charge can be precisely determined by the difference. In this way, standard DEPFET cells in circular 
geometry (see figure ~\ref{fig:DEPFET_CircularCell}) have been operated with an equivalent noise charge (ENC) of $4-5\,\mathrm{e^-}$ RMS for a readout time of $4\,\mathrm{\mu s}$~\cite{Treberspurg2016}. \\
The fact, however, that the quantity of charge is sensed indirectly via the channel conductivity enables an efficient implementation of a DEPFET device 
capable of RNDR. In case the charge is not cleared away during a CDS cycle, but transferred to an adjacent storage node, where the charge is still preserved and
where it has also no influence on the DEPFET channel conductivity, mimics a clear process and a nondestructive CDS cycle can be implemented. Transferring the charge 
back to the DEPFETs internal gate again after the CDS cycle has been finished starts a new CDS cycle for the identical signal charge. In case of the DEPFET, the 
second storage node can even be the internal gate of a second DEPFET adjacent to the first one, and the transfer can be conducted by means of an additional so-called 
transfer gate interposed between the two DEPFETs. The second DEPFET can also be used to conduct a CDS measurement, where the clear is replaced by the 
transfer back to the original DEPFET. This process can be repeated arbitrary times, 
until the charge is removed by the ClearFET after the final acquisition. 
In this way, one device pixel can be considered to be a {\it superpixel}  being composed of two DEPFET {\it subpixels}, whose internal gates are connected by 
the transfer gate. An example for a circuit representation and the respective layout is shown in figure~\ref{fig:DEPFET_RNDR}. 
\subsection{Performance model and prototype results}
\label{DEPFETPerformance}
\begin{figure}[]
\begin{center}
   \includegraphics[width=0.5\textwidth]{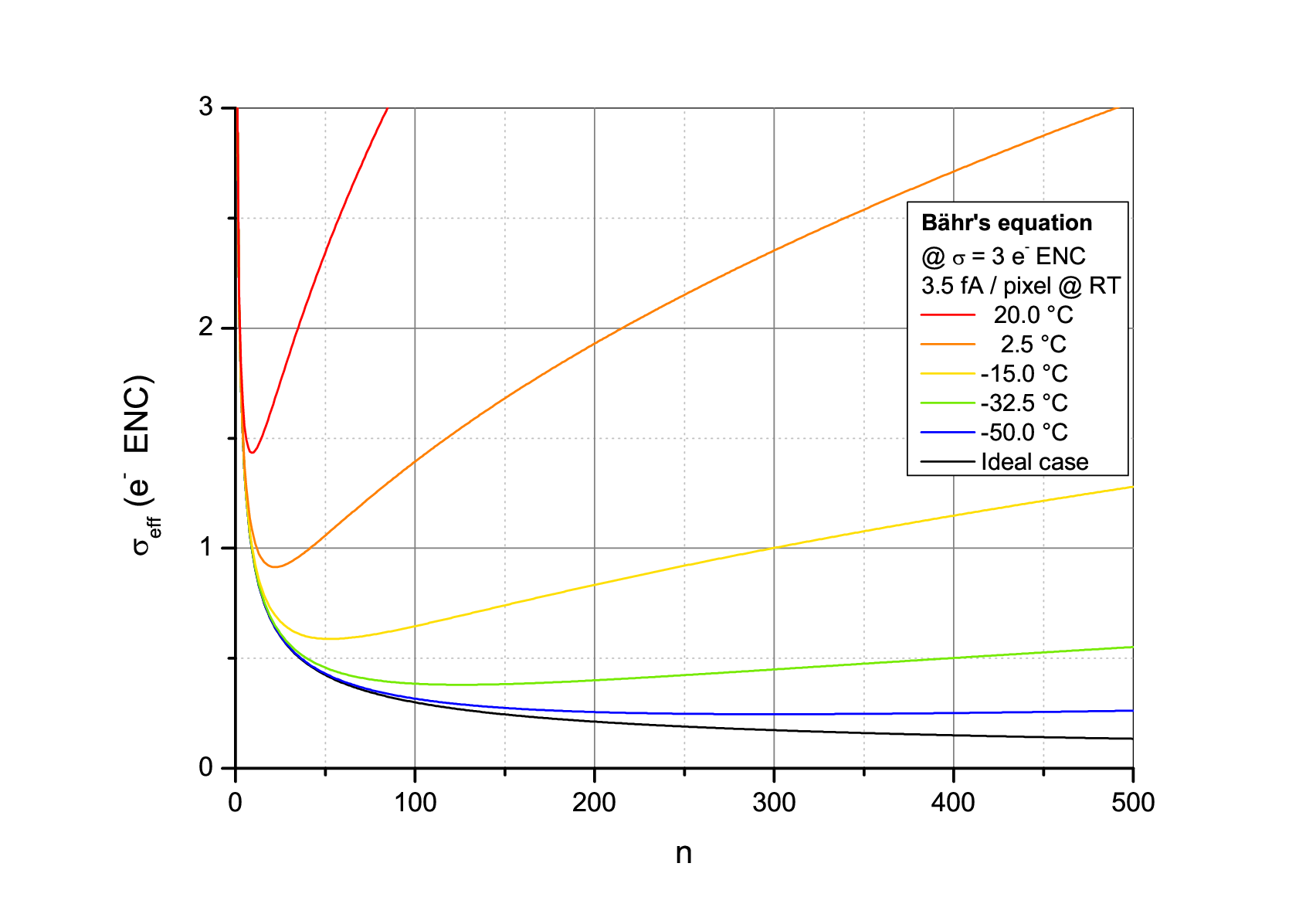}
  \includegraphics[width=0.5\textwidth]{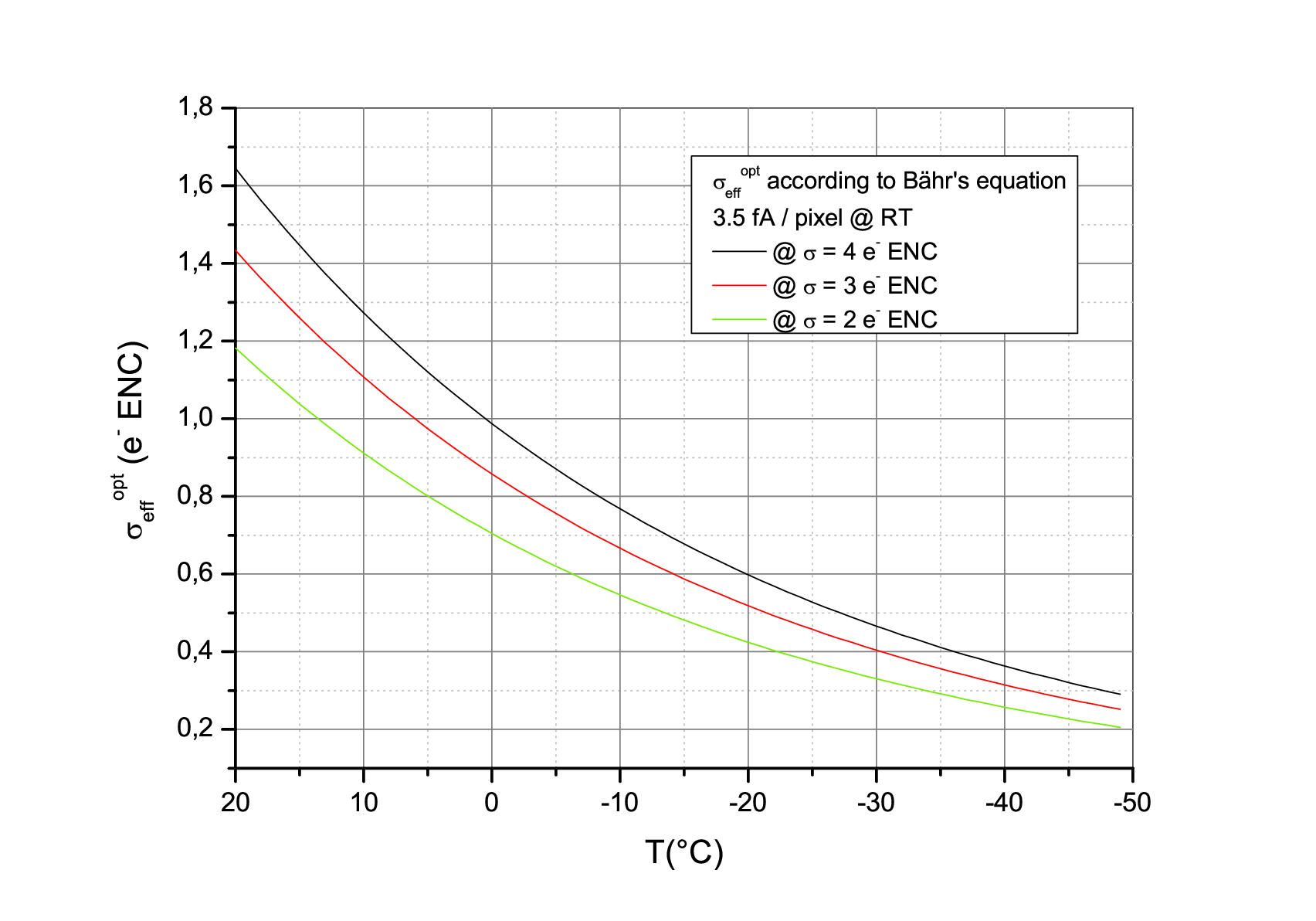}
\caption{Example performance graphs according to B\"ahr's equation: Dependence of $\sigma_\mathrm{eff}$ on $n$ and temperature for a pixel with $3\,e^{-}$ ENC 
(top) and dependence of $\sigma^\mathrm{opt}_\mathrm{eff}$ on temperature for three different initial sigma values.}
\label{fig:DEPFET_Performance}
\end{center}
\end{figure}

\begin{figure}[]
\begin{center}
   \includegraphics[width=0.5\textwidth]{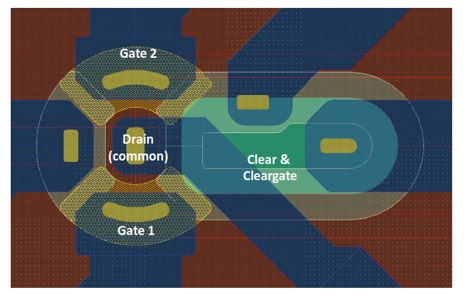}
  \includegraphics[width=0.5\textwidth]{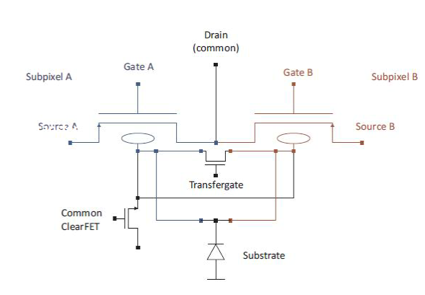}
\caption{Compact RNDR DEPFET {\it superpixel} layout in circular geometry as operated for the prototype tests (top) and equivalent circuit representation (bottom).}
\label{fig:Compact_DEPFET_RNDR}
\end{center}
\end{figure}

\begin{figure}[!htbp]
\begin{center}
   \includegraphics[width=0.45\textwidth]{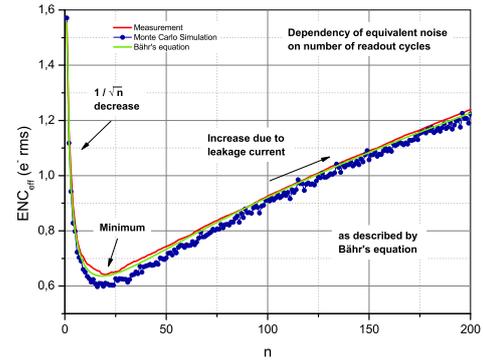}
   \includegraphics[width=0.45\textwidth]{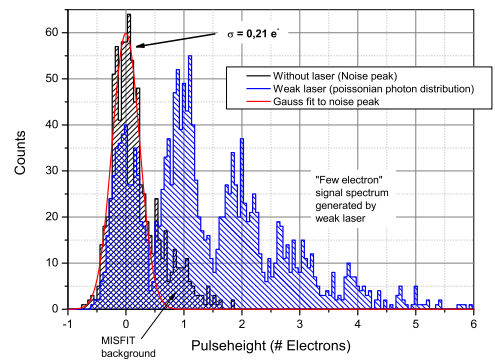}
    \includegraphics[width=0.45\textwidth]{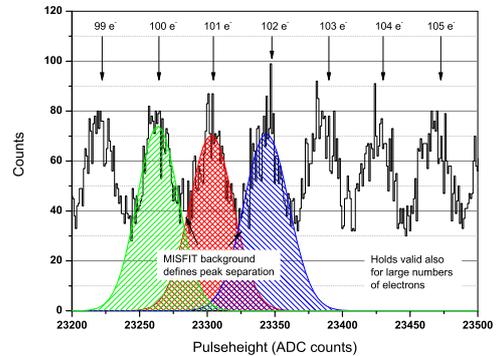}
\caption{Example of measurement results for RNDR DEPFET prototypes operated at $-40\,{}^{\circ}$C: agreement between measured 
$\sigma_\mathrm{eff}$ and $\sigma_\mathrm{eff}$ predicted both by Monte Carlo and B\"ahr's equation (top), single electron spectra taken for weak 
(middle) illumination intensities exhibiting the expected poisson distribution, and peak separation for higher illumination intensities 
(bottom). From the distance of the single electron peaks, a $\sigma^\mathrm{opt}_\mathrm{eff}$ of $0.2\,e^{-}$ ENC can be derived.}
\label{fig:DEPFET_RNDR_Performance}
\end{center}
\end{figure}

In practice, however, the RNDR process is disturbed by the advent of additional signal- and leakage electrons from the 
bulk during the signal evaluation process. 
This leads to a deviation from the ideal behaviour, which has been described by B\"ahr's equation~\cite{Baehr:Diplom}: 
\begin{equation}
\sigma_\mathrm{eff}^{2} = \frac{\sigma^{2}}{n}+\Delta \sigma^{2} \cdot
\left(\frac{1}{2}+\frac{1}{3}\cdot n - \frac{5}{6}\cdot \frac{1}{n} \right)
\end{equation}
where $\Delta \sigma$ is the expected increase in noise during one CDS acquisition in the RNDR cycle. For given  $\Delta \sigma$ and $\sigma$, an optimum number of transfer cycles can be derived:
\begin{equation}
n^\mathrm{opt}=\sqrt{3\cdot\frac{\sigma^{2}}{\Delta \sigma^{2}}-\frac{5}{2}}
\end{equation}
resulting in an optimum achievable effective noise of:
\begin{equation}
\sigma^\mathrm{opt}_\mathrm{eff} = \sqrt{\frac{\sigma^{2}}{n^\mathrm{opt}} +\Delta \sigma^{2}\cdot
\left(\frac{1}{2}+\frac{1}{3}\cdot n^\mathrm{opt} - \frac{5}{6}\cdot
\frac{1}{n^\mathrm{opt}}\right)}
\label{eq:sigma_opt}
\end{equation}
An example for the dependence can be seen in figure~\ref{fig:DEPFET_Performance}. 
The second summand under the square-root in Eq.~\ref{eq:sigma_opt} describes the deviation from the expected $1/\sqrt{n}$ behaviour due to 
the influence of the increase in noise $\Delta \sigma$ 
originating from the leakage current. Inside a 
silicon detector, this contribution can be efficiently suppressed, but not eliminated, by cooling. For low temperatures, the performance curve 
will approximate the ideal  $1/\sqrt{n}$ behaviour.  \\
A DEPFET based RNDR device optimised for the detection of the extremely weak signals (i.e. $\sigma < 2-3 \,e^{-}$ ENC) 
can be operated with an optimum number of readout cycles, which allows to lower 
$\sigma^\mathrm{opt}_\mathrm{eff}$ down to a level, where the minimum detectable signal (i.e. one electron) can be only generated by noise 
fluctuations with 5 sigma probability or lower. 
The application of cumulative measurement techniques (i.e. using the non-destructive readout 
without clearing of the pixel charge) helps to reduce this source of background 
(i.e. seeming single electron signals due to noise fluctuations) even further.  
Here, suppression of the $\Delta \sigma$-contribution in the perturbation term of
B\"ahr's equation to $10^{-4}$, and even lower, helps to achieve a $\sigma^\mathrm{opt}_\mathrm{eff}$ of $0.2\,e^-$ and below. 
This is achieved by adopting either the electronic shutter option or the Infinipix topology.
Nevertheless, even in case the detector is operated with an effective threshold of one electron,
volume leakage current collected during the sensors integration time is a source of 
irreducible background. \\
To maintain the sensitivity for the WIMP interaction signature as low 
as $2-3\,e^-$, the aim must be to lower the probability of two leakage 
current electrons within one pixel and
frame to as low a level as possible. This can be achieved by operating the device at lowest possible
temperatures to decrease the absolute magnitude of the leakage current, or by increasing the
readout rate to limit the integration time, or by a combination of both methods. Again, cumulative
measurements can help to preserve the statistical significance by preventing performance
deterioration due to recombination noise. \\
Standard mode RNDR DEPFET in circular geometry furnished with compact {\it subpixels} sharing the clear contact (see figure~\ref{fig:Compact_DEPFET_RNDR})
have been operated in single-pixel and small matrix environments for proof-of-principle measurements and verification of the performance model. Results have been 
reported in~\cite{Wolfel2007} and~\cite{Baehr:Diplom}, 
some results are shown in figure~\ref{fig:DEPFET_RNDR_Performance}. The predictions of B\"ahr's equation are nicely confirmed by both measurements 
and Monte Carlo simulations modelling the extended weighting 
function for the RNDR cycle. The value for  $\sigma_\mathrm{eff}^\mathrm{opt}$ of $0.18\,e^-$ RMS corresponds to the prediction for a device with a value for of $3\,e^-$ at 
$- 50\,{}^\circ $C and 256 transfer cycles. The single electron resolving capability was verified for amounts of charge of up to $10^3\,e^-$, the
peaks are nicely separated. \\
The high resistivity float zone silicon used for the fabrication of the sensors has a charge carrier lifetime at room temperature at the order of one second. This has to be seen in relation to the drift time in the depleting field, which, depending on the bias voltage, is at the order of 10 - $20\,\mathrm{ns}$. Cooling 
of the sensor increases the charge carrier lifetime to levels of minutes, so that an efficiency of $100\,\%$ for bulk generated electrons can be assumed with an accuracy of $10^{-9}$.
The  probability to create an electron-hole pair by the dark matter-electron scattering is fully described by Eq.~\ref{Eq:Scattering}. 
Highly doped regions on the front- and backside of the sensor, however, can be considered as dead material,
reducing the effective mass of the 
detector and therefore the exposure. The exposure quoted in the sensitivity studies described in section~\ref{ExpSensitivity} does not include the
dead material. The amount of dead material at sensor front- and backside, which is expected to be in the range of a few percent, 
needs to be determined by simulation using the final sensor layout. \\
\subsection{Planned improvements for future devices}
\label{DEPFETImproved}
\begin{figure}[!htbp]
\begin{center}
   \includegraphics[width=0.45\textwidth]{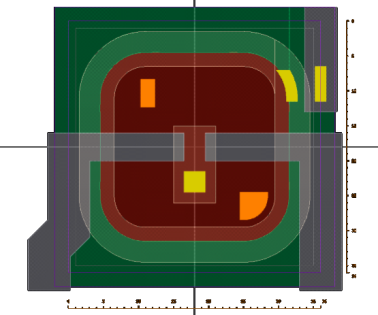}
  \includegraphics[width=0.450\textwidth]{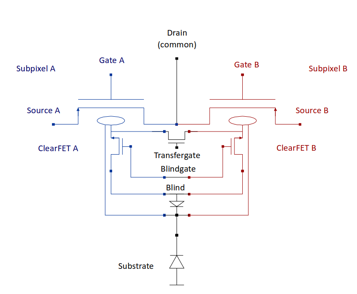}
\caption{
Layout (top) and equivalent circuit representation (bottom) of
RNDR {\it superpixel} with electronic shutter. The pixel size here is $36 \times 36\,\mathrm{\mu m^2}$.}
\label{fig:DEPFET_RNDR_Layout_2}
\end{center}
\end{figure}

In addition to the leakage current, a more serious perturbation of the RNDR process arises from the DEPFET's permanent sensitivity. In case signal charge arrives 
during the RNDR cycle, the signal charge is altered and the resulting mean value of the $n$ measurements does not represent the original signal charge. This is mainly a 
problem for applications were the incoming radiation is not synchronized with the readout cycle and for the background events for applications where it is. Although 
running average techniques can be applied during the RNDR process to detect the occurrence of these so-called misfit events, it is better to reduce their overall 
influence or even to completely avoid it. In this respect, two different approaches have been pursued to optimize RNDR-based detectors for future applications: 
\begin{itemize}
\item  A substantial reduction of the initial noise figure $\sigma$ for a single reading decreases not only $\sigma_\mathrm{eff}^\mathrm{opt}$ (see Eq.~\ref{eq:sigma_opt}), but also $n^\mathrm{opt}$ and, 
accordingly, the required time for the RNDR cycle. Constant signal rate provided, this in proportion reduces the probability for misfit events.
\item In addition, the introduction of a global electronic shutter to the pixel array decouples the DEFET {\it superpixels} from the detector bulk. 
Charge generated in the silicon bulk while the shutter is active will be extracted from the detector volume without being detected. 
Although this approach introduces some degree of dead-time, it provides a reduction of misfit background by at least two orders 
of magnitude in addition to the improvements achieved by reducing the noise.
\end{itemize}
Both options have been evaluated with respect to feasibility. 
Concerning the first option, an optimization of the DEPFET response by adapting geometry and
standard process technology parameters is expected to lower the initial noise figure down to values
of 2-$1.5\,e^{-}$ ENC, depending on the shaping time. This lowers both $n^\mathrm{opt}$ by an order of magnitude and,
accordingly, $\sigma^\mathrm{opt}_\mathrm{eff}$ to levels far below the single electron threshold~\cite{Treberspurg2017}.
More advanced modifications of the process technology, which are currently under investigation, 
have the potential to improve the performance even further. \\
The implementation of an electronic shutter has been evaluated via simulations and on a prototype level, and its functionality has been verified. 
The introduction of additional blind and blind-gate 
contacts surrounding the pixel structure allow to extract electrons on demand, providing a charge suppression factor of $10^{-3}$ and higher for the {\it superpixels}, while maintaining full 
retention of charge already stored in the internal gates. The shutter speed is below $100\,\mathrm{ns}$. Figure ~\ref{fig:DEPFET_RNDR_Layout_1} shows layout and circuit representation 
of a typical RNDR pixel with shutter functionality. \\
One of the biggest drawbacks of DEPFET based devices is the pixel size. Current RNDR DEPFET prototype devices exhibit pixel sizes at 
the order of $75 \times 75\,\mathrm{\mu m^2}$. This relatively large pixel size is partially 
counterbalanced by the full depletion in combination with the relatively large device thickness of $450\,\mathrm{\mu m}$. Nevertheless, this large pixel size 
limits the capability for background suppression on the base of cluster analysis 
especially for events in a very shallow depth beneath the pixel structure.  For this reason, compact devices have been designed, which provide 
for a pixel size of $36 \times 36\,\mathrm{\mu m^2}$. This very compact design (see 
figure~\ref{fig:DEPFET_RNDR_Layout_2}) has been realized by combining clear and shutter contacts. The design implements global clear and shutter functionality and allows for incremental 
as well as absolute charge 
measurements. In combination with the large bulk thickness, cluster analysis is possible to some extent. \\
For dark matter detection, arrays of 1k $\times$ 1k of these pixels are proposed covering an area of $\approx  3.7 \times  3.7\,\mathrm{cm^2}$, on a fully depleted detector 
bulk of $1\,\mathrm{mm}$ thickness. Detector mass is at the order of $3.2\,\mathrm{g}$. Initial noise is expected to be $1.5\,e^{-}$ ENC, target noise is $< 0.2\,e^{-}$ ENC.
\subsection{Planned prototype measurements}
\begin{figure}[!htbp]
\begin{center}
   \includegraphics[width=0.45\textwidth]{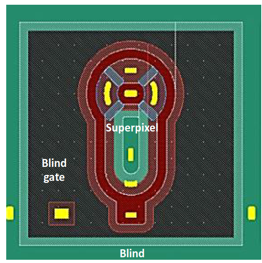}
  \includegraphics[width=0.450\textwidth]{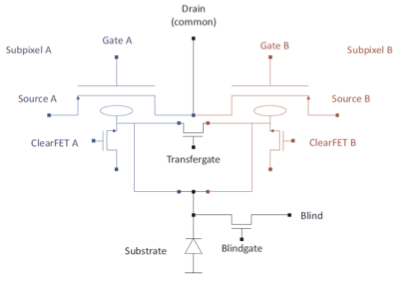}
\caption{Layout (top) and equivalent circuit representation (bottom) of
RNDR {\it superpixel} with electronic shutter. The pixel size here is $75 \times 75\,\mathrm{\mu m^2}$.}
\label{fig:DEPFET_RNDR_Layout_1}
\end{center}
\end{figure}

The base for the development of such devices will be the data gathered from the upcoming prototype measurements. Here, RNDR DEPFET devices with standard topology with and without global shutter functionality as shown in 
figure~\ref{fig:Compact_DEPFET_RNDR} and~\ref{fig:DEPFET_RNDR_Layout_1} respectively will for the first time be operated on a larger matrix scale in a low background environment.  
The devices consist of an array of 64 $\times$ 64 pixels integrated on a $0.45\,\mathrm{mm}$ thick silicon bulk. \\
Goal of the measurement is the complete parametrization of the devices in terms of operational parameters, operating temperature for lowest leakage current and optimized 
readout for optimum noise performance. The readout setup is optimized for background shielding and low noise rather than high speed readout, as the 
frame rate is at the order of mHz or even lower. 
\section{Dark Matter sensitivity studies}
\label{ExpSensitivity}
The low-noise measurement performance of the ionisation signal from an inelastic dark matter-electron scattering
measured with the RNDR DEPFET sensor described in section~\ref{DEPFET}
can be translated into a sensitivity for the detection of MeV dark matter. Like in section~\ref{SiliconDarkMatterDetection} 
we use the publicly available QEdark code~\cite{Essig:2015cda} for the estimate. \\ 
We analyse the impact of three key-parameters on the experimental sensitivity to low mass dark matter. 
Besides the threshold for the ionisation measurement, we study the exposure and the impact of 
background events on the sensitivity. While the DEPFET devices described 
in section~\ref{DEPFETImproved} have a mass of $3.1\,\mathrm{g}$ only, we will discuss
our results with a default exposure of one kg$\cdot$y and presents results with $0.1\,\mathrm{kg \cdot y}$ as
an alternative scenario. We investigate two main background  
sources, which could influence the sensitivity: background events caused by the
energy depositions from radioactive decays from inside or outside of the experiment and background events generated 
by the leakage current \linebreak present during the operation of the silicon sensor. The two background sources
have a different impact on the operation of the DEPFET device.
\subsection{Background events from the leakage current}
For the operation of the RNDR DEPFET detector a bias voltage is applied to the sensor. A very small
leakage current is generated in the sensor, which can lead to the collection of electrons in the 
internal gate. These
electrons from the leakage current generate background events. The size of the leakage current,
and therefore the number of background events, can be reduced by operating the device at lower 
temperatures. \\ 
Even the smallest known leakage current in silicon devices generate a significant 
background event rate for single pixel hits. The total number of background events 
from single electron events originating from the leakage current grows proportional to the total exposure time.
Any increase of the readout rate of the device will not change the picture. 
The situation changes for background events with two electrons collected in 
a single pixel, assuming the 
probability to generate a single electron from the leakage current
is uncorrelated. The probability to collect two electrons from the leakage current 
in the same pixel is significantly reduced and, in addition, the increase of the 
readout rate with a regular clear of the internal gates will further reduce 
the probability to collect two electrons from leakage current in the same pixel. 
Alternatively, RNDR DEPFET devices allow for cumulative measurements, as the charge within 
one {\it superpixel} does not necessarily have to be cleared, but may remain within the {\it superpixel} 
for later comparative measurements. This can help to detect the presence of leakage current electrons 
within a pixel  during a  ``reference'' acquisition, whose presence may be confirmed or disproved during 
subsequent reference acquisitions and can later be subtracted from the ``final'' acquisition data, thus 
combining the benefits of a fast readout rate without the drawback of increased noise hit rate. 
This feature, however, is mainly interesting in the case of relatively high initial noise values. \\
For this sensitivity 
studies we assume a default threshold of $Q=2\,e^{-}$; in addition we
also study the expected sensitivity for a threshold of $Q=1\,e^{-}$ and $Q=3\,e^{-}$.
The impact of background events from the leakage current is crucial and is 
subject to detailed device studies planned for the future.
\subsection{Simulation of background contributions from intrinsic radioactivity}
\label{ExpBG}
\begin{figure}[]
\begin{center}
\includegraphics[width=0.5\textwidth]{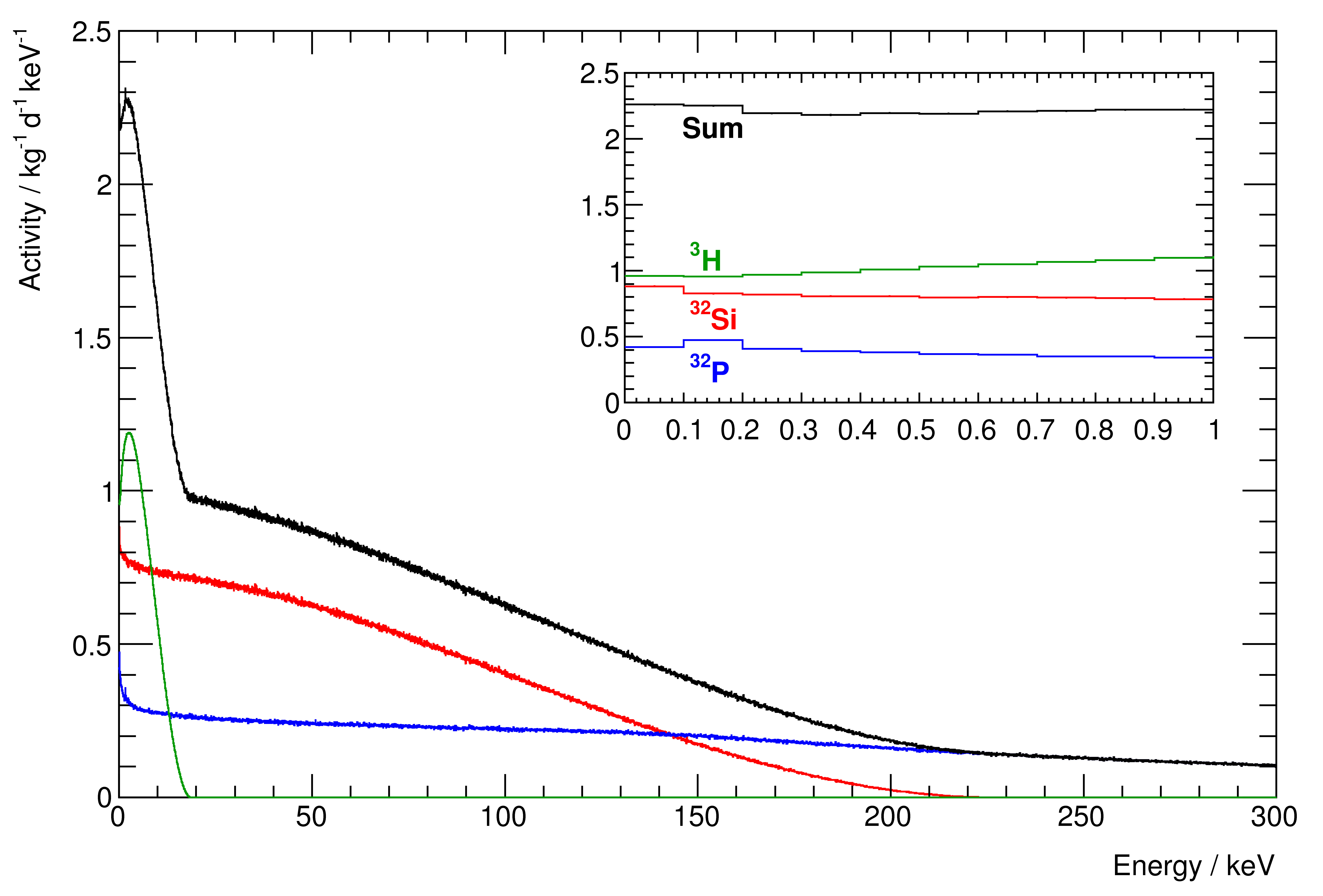}
\caption{Simulation of the energy deposition in a silicon sensor from
radioactive $^{32}$Si/$^{32}$P decays with a given activity of
$80\,\mathrm{kg^{-1}d^{-1}}$ and $^3$H decays with a given activity of $11.57\,\mathrm{kg^{-1}d^{-1}}$. The histograms show the energy depositions
from:
the $^{32}$Si decays (red), the subsequent $^{32}$P decay (blue), the $^3$H decay (green), and the sum of all 
decays (black). The inset zooms to the flat part of the spectrum below $1\,\mathrm{keV}$. Statistical uncertainties are comparable to the line width.}
\label{fig:SimulSi32}
\end{center}
\end{figure}

In~\cite{Essig:2015cda} a limit is derived for a background free experiment, while for this 
study we will discuss in addition the influence of background on the sensitivity of the experiment.
Background from radioactive decays can be subdivided in two different categories, intrinsic background
and background from external sources. We assume the shielding from external background
sources to be very efficient so that remaining external backgrounds
create surface events which can be rejected to a large extend, similar to the procedure 
used for detecting dark matter with semiconductor devices~\cite{Aguilar-Arevalo:2016ndq}. \\
An irreducible background from internal radioactive decays is expected.
The sensitive  detector elements consist mainly of silicon and previous studies
indicate, that the decay of $^{32}$Si is expected to be the leading contribution
to the internal background~\cite{Aguilar-Arevalo:2015lvd}.
Cosmogenic activation of Si can produce the unstable isotope $^{32}$Si, which
decays via $\beta^{-}$-decay with a half-life of $t_{1/2}=153\,\mathrm{y}$
and an energy release of $227.2\,\mathrm{keV}$. The decay leads to an energy
deposition in the sensor and generates background events in the region of interest.
The decay product $^{32}$P is also unstable and decays with a half-life of
$t_{1/2}=14.268\,\mathrm{d}$ and an energy release of $1.711\,\mathrm{MeV}$
to the stable isotope $^{32}$S. A further cosmogenic background is $^3$H produced via muon spallation
and inelastic scattering of neutrons on silicon \cite{Zhang:2016,Wei:2017}.
It undergoes a $\beta^{-}$-decay into the stable $^3$He 
with an energy release of $18.592\,\mathrm{keV}$.\\
We simulate the energy deposition in silicon of the $^{32}$Si and the
subsequent $^{32}$P decay with the GEANT4 simulation package in version 10.2p1 using mostly the default
processes described by the ``Low Energy Electromagnetic Physics Working
Group''~\cite{Allison:2016lfl,Allison:2006ve,Agostinelli:2002hh}. Only the size of the sampling bins of the
$\beta^-$ spectra are decreased by a factor 100 relative to the default settings to increase the precision at 
lowest energies.
We model a silicon only device with the geometry similar to the device to be
used for initial dark matter searches. We set the activity of $^{32}$Si in the
sensor to $80\,\mathrm{kg^{-1}d^{-1}}$~\cite{Aguilar-Arevalo:2015lvd}. The decaying
$^{32}$Si isotopes are randomly distributed in the sensor.
We explicitly note, since $^{32}$Si is generated via cosmogenic activation, that
the activity strongly depends on the time the silicon device is exposed to
cosmic rays and the activity might vary for other devices.\\
To our knowledge no measurement of the cosmogenic $^3$H production rate
$R_\mathrm{3H}$ in Si exists.
Therefore, we rely on the simulation study \cite{Zhang:2016} which found a
strong dependence of $R_\mathrm{3H}$ on the used simulation code, resulting in values ranging from 
27.29 to $108.74\,\mathrm{kg^{-1}d^{-1}}$. In a conservative approach we use the
upper limit and set $R_\mathrm{3H}=108.74\,\mathrm{kg^{-1}d^{-1}}$. The cosmogenic induced
activity $A_\mathrm{3H}$ is then given by \cite{Wei:2017}
\begin{equation}
A_\mathrm{3H}=R_\mathrm{3H}\cdot \left( 1-e^\frac{-\ln 2 \cdot t_\mathrm{exp}}{t_{1/2}} \right) \cdot e^\frac{-\ln 2 \cdot t_\mathrm{cool}}{t_{1/2}},
\end{equation}
where $t_{1/2}=12.32\,\mathrm{y}$ is the half-life of $^3$H, $t_\mathrm{exp}$
is the period of exposure to cosmic rays, and $t_\mathrm{cool}$ is the cooling
time, i.e.\ the duration at an underground location.
Assuming a duration of $t_\mathrm{exp}=2\,\mathrm{y}$ between growth of the Si
crystal and movement of the assembled detector to an underground laboratory, and
afterwards an immediate start of operation, i.e.\
$t_\mathrm{cool}=0\,\mathrm{y}$, results in
$A_\mathrm{3H}=11.57\,\mathrm{kg^{-1}d^{-1}}$. \\
The simulated spectrum of the deposited energy in silicon from $^{32}$Si,
$^{32}$P, and $^3$H decays is shown in figure~\ref{fig:SimulSi32}.
The simulation returns a flat spectrum with an activity of $\leq
2.26\,\allowbreak \mathrm{kg^{-1} d^{-1} keV^{-1}}$,  corresponding to 
$0.825\,\allowbreak \mathrm{kg^{-1} y^{-1} eV^{-1}}$, for energy depositions below
$1\,\mathrm{keV}$.\\
The RNDR DEPFET detector is able to detect single electrons with a resolution of $0.2\,e^{-}$. 
To estimate the total background rate we follow the conversion from the total deposited energy 
to ionization as used in~\cite{Essig:2015cda} and summarised in Eq.~\ref{AvIonisation}. 
We define as the signal region the energy range between the band gap energy of silicon ($1.1\,\mathrm{eV}$) 
and the minimum energy needed to generate three electrons ($8.3\,\mathrm{eV}$), corresponding to the
first two bins in figure.~\ref{fig:ER_Si}. By defining the first $Q$-bin as part of the signal region 
we follow a conservative approach and allow upward fluctuations of $Q=1$ to $Q=2$ hits,
generated by the leakage current.
With the given background activity of $0.825\,\mathrm{kg^{-1} y^{-1} eV^{-1}}$
for energy depositions below $1\,\mathrm{keV}$, as reported above, we 
expect a background rate of $5.94\, \mathrm{kg^{-1} y^{-1}}$ in the region of $Q=1$ to $Q=2$. 
For the  sensitivity studies we use a background rate of $6\, \mathrm{kg^{-1} y^{-1}}$.
\subsection{Expected sensitivity for detecting MeV dark matter with DETFET-RNDR detectors} 
\begin{figure}[!htbp]
\begin{center}
        \includegraphics[width=0.50\textwidth]{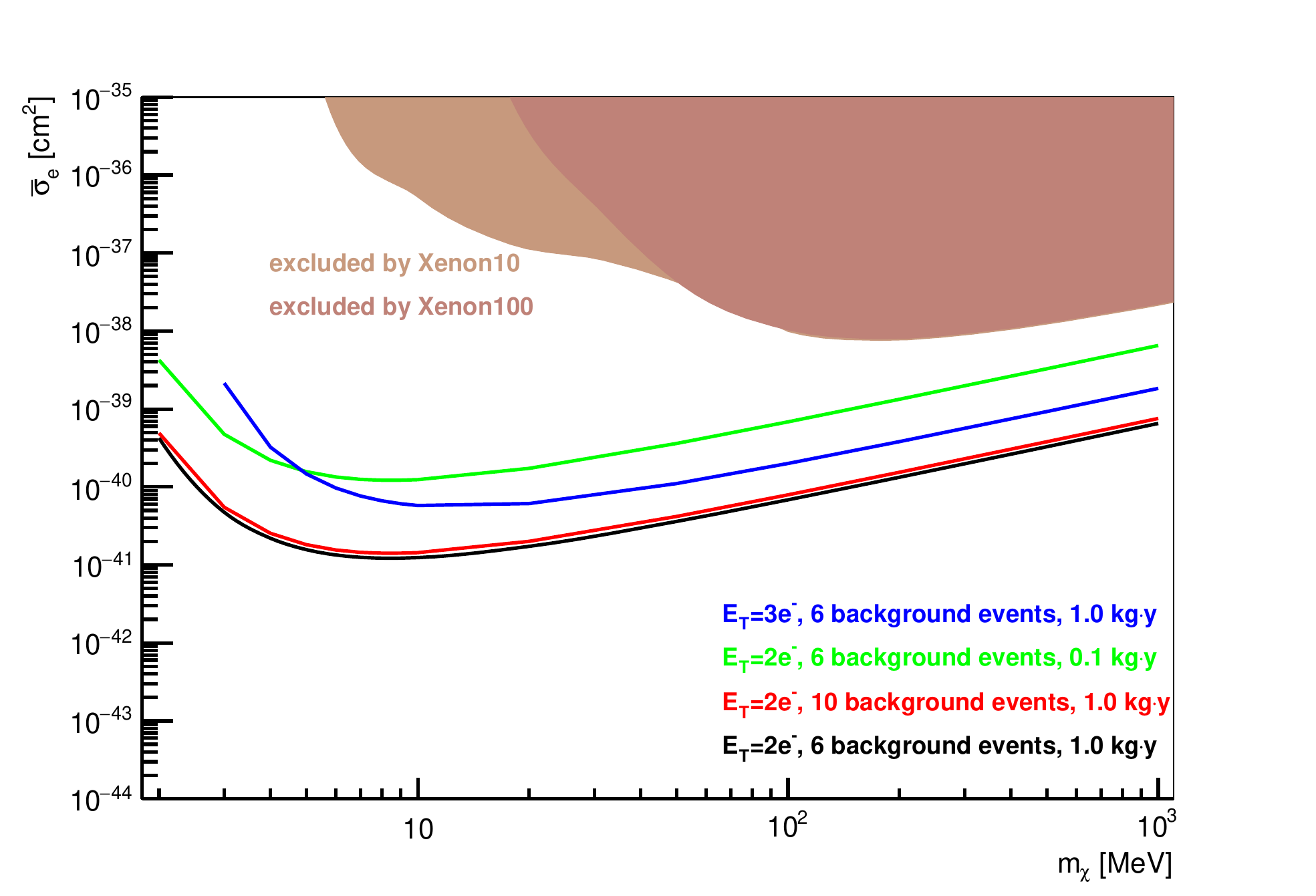}        
 \caption{Expected sensitivity for dark matter-electron scattering assuming a
threshold of $Q=2\,e^{-}$, six background events and an exposure of 
one kg$\cdot$y (black line). In addition we show the sensitivity 
with an increased threshold of $Q=3\,e^{-}$ (blue), reduced exposure of $0.1\,
\mathrm{kg \cdot y}$ (green) and increased background of ten events (red). We assume a constant form factor of $F_\mathrm{DM}(q)=1$.
For comparison the best limit using data from Xenon10 and Xenon100 is shown ~\cite{Essig:2017kqs}.}
\label{fig:BgAssumption}
\end{center}
\end{figure}

\begin{figure}[!htbp]
\begin{center}
        \includegraphics[width=0.50\textwidth]{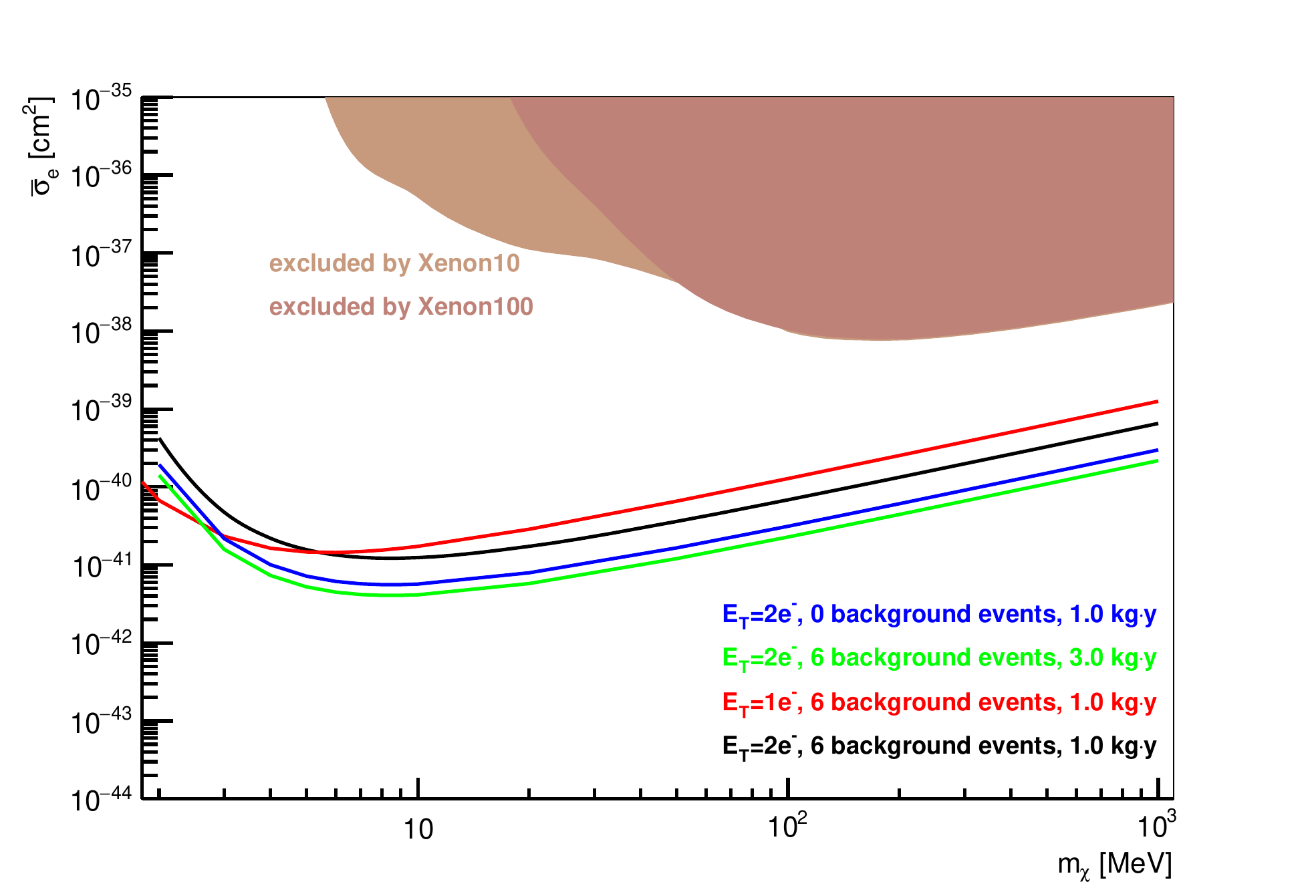}
\caption{Expected sensitivity for dark matter-electron scattering assuming a
threshold of $Q=2\,e^{-}$, six background events and an exposure of 
one kg$\cdot$y (black line). In addition we show the sensitivity 
with a reduced threshold of $Q=1\,e^{-}$ (red), increased exposure of $3\,
\mathrm{kg \cdot y}$ (green) and decreased background of zero events (blue).  We assume a constant form factor of $F_\mathrm{DM}(q)=1$.
For comparison the best limit using data from Xenon10 and Xenon100 is shown ~\cite{Essig:2017kqs}.}
\label{fig:Threshold}
\end{center}
\end{figure}

We use the number of predicted background events from $^{32}$Si,
$^{32}$P, and $^3$H decays together with
the code QEdark code to calculate the expected sensitivity of the experiment~\cite{Essig:2015cda}. 
We consider a constant form factor of $F_\mathrm{DM}(q)=1$ only.
We determine the expected sensitivity assuming six background events, an exposure of one
kg$\cdot$y and a threshold of $Q=2\,e^{-}$. We use the statistical approach
described in~\cite{Feldman:1997qc} to determine the expected sensitivity. We assign no 
uncertainty to the number of expected background events and we take the number
of observed events to be equal to the number of background events. The upper 
limit for the number of signal events for six background events is 6.75 events (95 $\%$ C.L.). \\
The expected sensitivity for different assumptions is shown in figure~\ref{fig:BgAssumption} and
figure~\ref{fig:Threshold}. Please note, that we assume in all cases no background events from 
leakage current events. With the default assumption of an energy threshold
of two electrons, six background events and an exposure of one kg$\cdot$y 
we can reach a sensitivity of about $\overline{\sigma}_\mathrm{e} = 10^{-41}\,\mathrm{cm^2}$ for 
dark matter particles with a mass of $10\,\mathrm{MeV}$. Assuming six background 
events the maximal sensitivity can be reached with an exposure of about one kg$\cdot$y, an
exposure of three years improves the sensitivity only marginally. Increasing the
threshold from two electrons to three electrons reduces the sensitivity in the 
MeV mass region by almost one order of magnitude. A reduction
of the exposure to $0.1\,\mathrm{kg \cdot y}$ will lead to a similar loss in sensitivity.

\section{Conclusion}
\label{Summary}
The quest for particle dark matter is among the most urgent open topics of modern physics. The 
mass range for dark matter candidates, as well as the interaction rate with ordinary matter, is
unpredicted. 
The parameter space for light dark matter in the MeV mass range 
still has some experimentally unexplored regions. We discuss the possibility to 
use a silicon detector operated as a RNDR DEPFET device to detect single electrons 
being produced by possible dark matter-electron scatterings.  
Measurements using a RNDR DEPFET prototype
return an effective noise of $0.18\,e^{-}$ RMS, allowing to resolve single electrons. 
Assuming six \linebreak background events in the signal region, a threshold of two electrons and an exposure of one kg$\cdot$y, we determine 
the expected sensitivity to be about $\overline{\sigma}_\mathrm{e} = 10^{-41}\,\mathrm{cm^2}$ for dark matter particles with a mass of $10\,\mathrm{MeV}$.

% Non-BibTeX users please use


\begin{thebibliography}{}
\bibitem{DMVelocity}
The dark matter density is set to $0.4\,\mathrm{GeV/cm^3}$, a Maxwell-Boltzmann
velocity distribution for the dark matter particles is assumed with $300\,\mathrm{km/s}$ as 
mean velocity and the escape velocity is 
set to $600\,\mathrm{km/s}$. % and the average earth velocity is 240 km/s.
\end{thebibliography}
\end{document}